\documentstyle[11pt,epsf]{article}

\newcommand{\RR}{\mbox{${\rm \:  R\!\!\!\! I
\;\;}$}}

\newcommand{\vs}{\vspace{0.25cm}}

\newtheorem{theorem}{Theorem}
\newtheorem{itlemma}{Lemma}[section]
\newtheorem{itproposition}[itlemma]{Proposition}
\newtheorem{itcorollary}[itlemma]{Corollary}
\newtheorem{itremark}[itlemma]{Remark}
\newtheorem{itremarks}[itlemma]{Remarks}
\newtheorem{itdefinition}[itlemma]{Definition}
\newtheorem{itexample}[itlemma]{Example}

\newenvironment{lemma}{\begin{itlemma}\rm}{\end{itlemma}} 
\newenvironment{remark}{\begin{itremark}\rm}{\end{itremark}} 
\newenvironment{remarks}{\begin{itremarks} \rm}{\end{itremarks}}
\newenvironment{corollary}{\begin{itcorollary}\rm}{\end{itcorollary}}
\newenvironment{proposition}{\begin{itproposition}\rm}{\end{itproposition}}
\newenvironment{definition}{\begin{itdefinition}\rm}{\end{itdefinition}}
\newenvironment{example}{\begin{itexample}\rm}{\end{itexample}}
\newenvironment{fact}{\noindent {\em Fact}. \ \ }{\hfill \medskip}
\newenvironment{proof}{\noindent {\em Proof}.\ \
}{\hspace*{\fill}$\Box$\medskip}
\newenvironment{claim}{\noindent {\em Claim}. \ \ }{\hfill \medskip}
\newcommand{\be}[1]{\begin{equation}\label{#1}}
\newcommand{\ee}{\end{equation}}
\newcommand{\bl}[1]{\begin{lemma}\label{#1}}
\newcommand{\br}[1]{\begin{remark}\label{#1}}
\newcommand{\brs}[1]{\begin{remarks}\label{#1}}
\newcommand{\bt}[1]{\begin{theorem}\label{#1}}
\newcommand{\bd}[1]{\begin{definition}\label{#1}}
\newcommand{\bp}[1]{\begin{proposition}\label{#1}}
\newcommand{\bc}[1]{\begin{corollary}\label{#1}}
\newcommand{\bfact}[1]{\begin{fact}\label{#1}}
\newcommand{\bex}[1]{\begin{example}\label{#1}}
\newcommand{\ec}{\end{corollary}}
\newcommand{\efact}{\end{fact}}
\newcommand{\eex}{\end{example}}
\newcommand{\el}{\end{lemma}}
\newcommand{\er}{\end{remark}}
\newcommand{\ers}{\end{remarks}}
\newcommand{\et}{\end{theorem}}
\newcommand{\ed}{\end{definition}}
\newcommand{\ep}{\end{proposition}}
\newcommand{\epr}{\end{proof}}
\newcommand{\bpr}{\begin{proof}}
\newcommand{\bcl}{\begin{claim}}
\newcommand{\ecl}{\end{claim}}

\newcommand{\bi}{\begin{itemize}}
\newcommand{\ei}{\end{itemize}}
\newcommand{\ben}{\begin{enumerate}}
\newcommand{\een}{\end{enumerate}}
\newcommand{\text}[1]{\hbox{\rm \ #1\ \/}}

\begin{document}

\begin{center}

{\Large Quantum Symmetries and Cartan Decompositions in Arbitrary
Dimensions}

\vs


Domenico D'Alessandro\footnote{Department of Mathematics, Iowa
State University, Ames,  IA 50011, U.S.A., Tel.:+1-515-2948130,
Fax: +1-515-2945454, e-mail: daless@iastate.edu } and  Francesca
Albertini\footnote{Department of Pure and Applied Mathematics,
University of Padova, Via Belzoni 7, 35100 Padova, Italy, Tel.:
+39-49-8275966, Fax: +39-49-8275892,
e-mail:albertin@math.unipd.it}

\end{center}

\begin{abstract}

We investigate the relation between Cartan decompositions of the
unitary group and discrete quantum symmetries. To every Cartan
decomposition there corresponds a quantum symmetry which is the
identity when applied twice. As an application, we describe a new
and general method to obtain Cartan decompositions of the unitary
group of evolutions of multipartite systems from Cartan
decompositions on the single subsystems. The resulting
decomposition, which we call of the {\it odd-even type}, contains,
as a special case,  the {\it concurrence canonical decomposition}
(CCD) presented in \cite{BBO1},\cite{BBO2},\cite{BDN} in the
context of entanglement theory. The CCD is therefore extended from
the case of a multipartite system of $n$ qubits to the case where
the component subsystems have arbitrary dimension.

\end{abstract}













\section{Introduction}

Decompositions of Lie groups have been extensively used in control
theory to design control algorithms for bilinear, right invariant,
systems with state varying on a Lie group. Once it is known how to
factorize a target final state $X_f$ as the product \be{factoriz}
X_f=X_1 X_2 \cdot \cdot \cdot X_r, \ee then the task of
controlling to $X_f$ can be reduced to the (simpler) task of
controlling to the factors $X_1,...,X_r$. In quantum information
theory, a factorization of the type (\ref{factoriz}) can be
interpreted as the implementation of a quantum logic operation
with a sequence of elementary operations. In this case,  the
relevant Lie group is the Lie group of unitary matrices of
dimensions $n$, $U(n)$. In general,  a decomposition of the
unitary evolution operator of the form (\ref{factoriz}) is useful
to determine several aspects of the dynamics of quantum systems
including the degree of entanglement (see e.g. \cite{Zhang}), time
optimality of the evolution \cite{KBG} and constructive
controllability (see e.g. \cite{M1}, \cite{M3}).

Most of the studies presented so far, which involve  Lie group
decompositions applied to the quantum systems, are concerned with
low dimensional systems. For these systems, several   complete and
elegant results can be obtained, which also have important
physical implications. Decompositions of the unitary group $U(n)$
for large $n$ exist and can be applied to the dynamical analysis
of high dimensional quantum systems. However, the information
obtained with this study is rarely as useful and of direct
physical interpretation as in the low dimensional cases. For
multipartite systems,  this motivates the search for Lie group
decompositions constructed in terms of decompositions on the
single subsystems. We shall construct such type of  decomposition
in the present paper.

The main motivation for the study presented here was given by the
recent papers \cite{BBO1} \cite{BBO2}, \cite{BDN}. In these
papers, a decomposition of $U(2^n)$ called the Concurrence
Canonical Decomposition was obtained for a quantum system of $n$
two level systems (qubits). Such a decomposition has the above
mentioned feature of being expressed in terms of elementary
decompositions on the single qubit subsystems. It is related to
time reversal symmetry and this raises the question of what in
general the relation is between quantum mechanical symmetries and
decompositions. As we shall see here, the answer to this
fundamental question is instrumental in developing a general
method to construct decompositions of multipartite systems from
elementary decompositions of the single subsystems. We shall
develop a decomposition which we call of the `odd-even type' that
contains the concurrence canonical decomposition as a special
case.

The paper is organized as follows. In Section \ref{BM} we review
the basic definitions and results concerning discrete quantum
symmetries and Cartan decompositions of the Lie algebra $su(n)$
and therefore the Lie group $SU(n)$. We shall stress the important
result that, up to conjugacies, there are only three types of
Cartan decompositions which are usually labeled as {\bf AI}, {\bf
AII} and {\bf AIII}. In Section \ref{RBC}, we investigate the
relation between Cartan decompositions and quantum symmetries and
establish a one to one correspondence between Cartan
decompositions and a subclass of symmetries which we call Cartan
symmetries. To every Cartan decomposition of the Lie algebra
$u(n)$ and corresponding Cartan symmetry there corresponds a
decomposition of the Jordan algebra of Hermitian matrices of
dimension $n$, $iu(n)$ equipped with the anticommutator operation.
This is described in Section \ref{DSO}. This is also the crucial
fact used to develop the general decomposition of the odd-even
type for multipartite systems in arbitrary dimensions in Section
\ref{EMS}. This decomposition is a Cartan decomposition and, in
Section \ref{TNO}, we show how to determine its type ({\bf AI} or
{\bf AII}). The Cartan decomposition also leads to a decomposition
of the evolution of any quantum system into the product of an
evolution with antisymmetric Hamiltonian and one with symmetric
Hamiltonian with respect to a  Cartan symmetry. This result is
discussed in Remark \ref{DD}.

\section{Background material}
\label{BM}

\subsection{Discrete Symmetries in
Quantum Mechanics} Given a quantum system with underlying  Hilbert
space ${\cal H}$,  a quantum mechanical {\it symmetry} is defined
(see e.g. \cite{GAPA} Chapter 7, \cite{Sakurai} Chapter 4) as a
one to one and onto map  $\Theta: {\cal H} \rightarrow {\cal H}$
such that physically indistinguishable states are also mapped into
physically indistinguishable states i.e. for every $|\psi> \in
{\cal H} $ and $\phi_1 \in \RR$, \be{requi} \Theta(e^{i \phi_1}
|\psi>)=e^{i \phi_2} \Theta (|\psi>), \ee for some $\phi_2 \in
\RR$. Moreover $\Theta$ preserves the inner product of two states,
namely if $|\tilde \alpha>:=\Theta |\alpha>$, then, for any two
states $|\alpha>$ and $|\beta>$ \be{inpp} |<\tilde \alpha | \tilde
\beta>|= |<\alpha | \beta>|. \ee In this definition, we omit for
simplicity the consideration of selection rules and assume that
all the states are physically realizable. A detailed discussion of
this point can be found in \cite{GAPA}.

According to Wigner's theorem \cite{Wigner}, every such operation
$\Theta$  can be represented as \be{Wigner}  \Theta=e^{i\phi}U,
\ee where $\phi$ is a constant, physically irrelevant, real
parameter, and $U$ is either a {\it unitary} operator or an {\it
anti-unitary} one. Recall that an anti-unitary operator $U$,
$|\alpha> \rightarrow |\tilde \alpha> :=U |\alpha>$ is defined as
satisfying \be{au1} <\tilde \beta | \tilde \alpha>= < \beta |
\alpha>^*,  \ee \be{au2} U ( c_1 | \alpha> + c_2 | \beta>) = c_1^*
U |\alpha>+ c_2^* U|\beta>.   \ee


Once a basis of the Hilbert space $\cal H$ is chosen, an
anti-unitary operator $U$ can always be written as \be{gottfried}
U| \alpha> = X K |\alpha>,  \ee where $K$ is the operation which
conjugates all the components of the vector $|\alpha>$ and $X$ is
unitary.


A symmetry $\Theta$, whether unitary or anti-unitary, induces a
transformation on the space of Hermitian operators $A$ as
\be{transA} A \rightarrow \Theta A \Theta^{-1}:=\bar \theta(A).
\ee It is in fact easily verified that $\bar \theta(A)$ is a
linear and Hermitian operator. Moreover the eigenvalues of $\bar
\theta(A)$ are the same as those of $A$ and a set of orthonormal
eigenvectors are given by $\Theta |\alpha_j>$, where $|\alpha_j>$
is an orthonormal basis of eigenvectors of $A$.
It can be proved \cite{GAPA}, \cite{Sakurai}, that, up to a phase
factor, $\bar \theta(A):=\Theta A \Theta^{-1}$ is the only choice
that guarantees \be{onlychoice} | < \tilde \alpha| \bar \theta(A)
| \tilde \beta> =|< \alpha| A | \beta>|.  \ee

Description of the symmetry $\Theta$ is usually done by specifying
 how $\bar \theta$ acts on Hermitian operators rather than how $\Theta$
 acts on states. This is because, Hermitian operators represent
 physical observables and therefore the action of $\theta$ on observables is
 typically suggested by physical considerations. For example, the
 {\it space translation symmetry} has to be such that
 \be{space}
\bar \theta(\hat x)=\hat x -a,
 \ee
for some constant $a$ where  $\hat x$ is the position operator. As
another example, the {\it parity} or  {\it space inversion
symmetry}
 is defined  such that \be{parity} \bar \theta (\hat x) = - \hat x.  \ee
On the other hand, specification of $\bar \theta$ on an
irreducible set of observables uniquely determines $\Theta$ up to
a phase factor \cite{GAPA}. Recall that an irreducible set of
observables $\{ A_j \}$ is defined such that if an observable $B$
commutes with all of the $\{ A_j \}$, then $B$ is a multiple of
the identity.



An observable $H$ is said to {\it satisfy a symmetry}  $\Theta$ or
{\it to be symmetric with respect to $\Theta$} if \be{symsym1}
\bar \theta(H)=H, \ee or equivalently \be{symsym2} \Theta H= H
\Theta. \ee It is said to be {\it antisymmetric with respect to
$\Theta$} if \be{symsym3} \bar \theta(H)=-H \leftrightarrow \Theta
H=-H \Theta. \ee


A special type of symmetry is the {\it time reversal symmetry}. In
classical mechanics a time reversal symmetry changes a system into
one which evolves with time reversal trajectories. This suggests
to define a time reversal symmetry in quantum mechanics so that
$\bar \theta$ acts on the position $\hat x$ and momentum operator
$\hat p$ according to \be{actonpos} \bar \theta(\hat x)=\hat x,
\ee
 \be{actonmom} \bar \theta(\hat p)=-\hat p.    \ee
This implies that the corresponding $\Theta$ transforms momentum
eigenvectors $|p>$ as \be{kj} \Theta |p> = |-p>. \ee  If the
system under consideration has no spin degree of freedom, then
$\hat x$ and $\hat p$ form  an irreducible set of observables and
therefore (\ref{actonpos}) and (\ref{actonmom}) uniquely specify
the transformation $\Theta$ on the state. Moreover, from the
definition of angular momentum $\hat L:= \hat x \times \hat p$, we
obtain \be{defianm} \bar \theta(\hat L)=\bar \theta(\hat x) \times
\bar \theta(\hat p)=- \hat L. \ee For a system with spin angular
momentum $\hat S$, we impose by definition, according to
(\ref{defianm}) \be{spinangmom} \bar \theta(\hat S)= -\hat S,  \ee
and $\hat x$, $\hat p$, $\hat S$ form an irreducible set of
observables. If $|m>$ is an eigenvector of the (spin) angular
momentum corresponding to eigenvalue $m$, we have \be{ool} \Theta
|m>=|-m>. \ee From these specifications, it is possible to obtain
an explicit expression of the time reversal symmetry for a system
of $N$ particles with spin operators $\hat S_1$,...,$\hat S_N$. It
is given (in a basis of tensor products of the eigenstates of the
$z-$ component of the spin operators) by (see \cite{GAPA}) \be{th}
\Theta=e^{-\frac{i \pi}{\hbar} (\hat S_{1,y}+\hat S_{2,y}+...+\hat
S_{N,y})} K,  \ee where $\hat S_{j,y}$ is the $y$ component of the
spin operator corresponding to the $j-$th particle, $j=1,...,N$,
and $K$ is the conjugation operator (same as in
(\ref{gottfried})).

\subsection{Cartan involutions and decompositions of $su(n)$}

We discuss next the Cartan decompositions for the Lie algebra
$su(n)$. What we say  could be generalized to general semisimple
Lie algebras. We refer to \cite{Helgason} for more details.

A {\em{Cartan decomposition}} of $su(n)$ is a vector space
decomposition \be{Cart} su(n)={\cal K} \oplus {\cal P}, \ee where
the subspaces $\cal K$ and $\cal P$ satisfy the commutation
relations

\be{comm1} [{\cal K}, {\cal K}] \subseteq {\cal K},  \ee
\be{comm2}[{\cal K}, {\cal P}] \subseteq {\cal P},  \ee \be{comm3}
[{\cal P}, {\cal P}] \subseteq {\cal K}. \ee In particular, notice
that $\cal K$ is a subalgebra of $su(n)$. A Cartan decomposition
of $su(n)$ induces a factorization of the elements of the Lie
groups $SU(n)$. Let us denote by $e^{\cal L}$ the connected Lie
group associated to a generic Lie algebra $\cal L$. Then, given a
Cartan decomposition  (\ref{Cart}), every element $X$ in $SU(n)$
can be written as \be{elem} X=KP,  \ee where $K \in e^{\cal K}$
and $P$ is the exponential of an element of  ${\cal P}$. Moreover
if $\cal A$ is a maximal Abelian subalgebra of $su(n)$, with
${\cal A} \subseteq {\cal P}$, then one can prove that \be{union2}
\cup_{K \in e^{\cal K}}K{\cal A}K^*= {\cal P}.   \ee This implies
that one can write $P$ in (\ref{elem}) as $P=K_1 A K_1^*$ with
$K_1 \in e^{\cal K}$ and $A \in e^{\cal A}$. Therefore every
element $X$ in $SU(n)$ can be written as \be{KAK} X=K_1AK_2,  \ee
with $K_1,K_2 \in e^{\cal K}$ and $A \in e^{\cal A}$. This is
often referred to as {\it $KAK$ decomposition}.


A {\it Cartan involution} of $su(n)$ is a homomorphism
 $\theta: su(n) \rightarrow su(n)$ such that $\theta^2$
 is equal to the identity on $su(n)$.  Associated to a
Cartan decomposition (\ref{Cart}) is a Cartan involution which is
equal to the identity on $\cal K$ and multiplies by $-1$ the
elements of $\cal P$, i.e. \be{invonK} \theta(K)=K, \qquad \forall
K \in {\cal K}, \ee \be{invonP} \theta(P)=-P, \qquad \forall P \in
{\cal P}. \ee Therefore, given a Cartan decomposition, relations
(\ref{invonK}) and (\ref{invonP}) determine a Cartan involution
$\theta$. Viceversa given a Cartan involution $\theta$,  the $+1$
and $-1$ eigenspaces of $\theta$ determine a Cartan decomposition.


According to a theorem of Cartan \cite{Helgason}, there exist only
three types of Cartan decompositions for $su(n)$ up to conjugacy.
More specifically, given a Cartan decomposition (\ref{Cart}) there
exists a matrix $H \in SU(n)$ such that ${\cal K}':=H{\cal K}
H^*$, ${\cal P}':=H{\cal P} H^*$, where ${\cal K}'$ and ${\cal
P}'$ fall in one of the following cases labeled {\bf AI}, {\bf
AII} and {\bf AIII}\footnote{In the following definitions and in
the rest of the paper the inner product $<A,B>$ in $su(n)$ is
defined as $<A,B>:=Tr(AB^*)$ and it is proportional to the Killing
form (see e.g. \cite{Helgason})} .


{\bf AI} \be{AIdec} {\cal K}'=so(n), \quad {\cal P}'=so(n)^\perp,
\ee where $so(n)$ is the Lie algebra of real
 skew-Hermitian matrices of dimension $n$,
 $so(n)^\perp$ is the vector space over the reals of purely
imaginary skew-Hermitian matrices. The corresponding Cartan
involution, which we denote by $\theta_I$, returns the complex
conjugate of a matrix, i.e. \be{AIinvo} \theta_I(A):=\bar A. \ee


{\bf AII} \be{AIIdec} {\cal K}'=sp(\frac{n}{2}), \quad {\cal
P}'=sp(\frac{n}{2})^\perp,  \ee where, we are assuming $n$ even,
and $sp(\frac{n}{2})$ is the Lie algebra of symplectic $n \times
n$ matrices i.e. the Lie algebra of skew-Hermitian matrices $A$
satisfying \be{symplcond} AJ+JA^T=0.  \ee The matrix $J$ is
defined as \be{matJ} J:=\pmatrix{0 & I_{\frac{n}{2}} \cr
-I_{\frac{n}{2}} & 0}.  \ee The corresponding Cartan involution
$\theta_{II}$ is given by \be{AIIinvo}\theta_{II}(A):=J \bar A
J^{-1}=-J \bar A J. \ee


{\bf AIII}. In this case ${\cal K}'$ is the set of all the
skew-Hermitian matrices $A$ of the form \be{formA} A=\pmatrix{R &
0 \cr 0 & S},  \ee where $R \in u(p),$ $S \in u(q)$, $p,q >0$,
$p+q=n$ and $Tr(R)+Tr(S)=0$. ${\cal P}'$ is equal to ${{\cal
K}'}^\perp$. The corresponding Cartan involution is given by
\be{AIIIinvo}\theta_{III}(A):= I_{p,q} A I_{p,q}, \ee where the
matrix $I_{p,q}$ is defined as the block matrix
$I_{p,q}:=\pmatrix{I_{p \times p} & 0 \cr 0 & -I_{q \times q}}$.


Several authors have proposed Lie algebra decompositions for
$su(n)$ that,  although special cases of the general Cartan
decomposition, are of particular significance in some contexts.
For example, Khaneja and Glaser \cite{KG} (see also \cite{Bullock}
for the relation of this decomposition with Cartan decomposition)
have factorized unitary evolutions in $SU(2^n)$, namely unitary
evolution of $n$ two level quantum systems ({\it qubits}), into
local operations i.e. operations on only one qubit and two-qubits
operations. This result has consequences both in the study of
universality of quantum logic gates and in control theory. In the
latter context,
 one would like to decompose the task of steering the evolution operator
to a prescribed target into a sequence of steering problems to
intermediate targets with a determined structure.

Another decomposition which is of particular interest to us is the
{\it Concurrence Canonical Decomposition} (CCD) of $su(2^n)$ which
was studied in \cite{BBO1}, \cite{BBO2}, \cite{BDN} in the context
of {\it entanglement} and {\it entanglement dynamics}. In this
decomposition, ${\cal K}'$ and ${\cal P}'$ are  real span of
tensor products, multiplied by $i$, of $n$ $2 \times 2$ matrices
chosen in the set $\{I_{2 \times 2}, \sigma_x,\sigma_y,
\sigma_z\}$, where $\sigma_{x,y,z}$ are the $x,y,z$ Pauli
matrices. In particular, ${\cal K}'$ is spanned by tensor products
with an {\it odd} number of Pauli matrices and ${\cal P}'$ is
spanned by tensor products with an {\it even} number of Pauli
matrices. It was shown in \cite{BBO1}, \cite{BBO2} that for $n$
even this decomposition is a Cartan {\bf AI} decomposition and for
$n$ odd is a Cartan {\bf AII} decomposition. One of the primary
goals of the present paper is to extend the CCD to the case of
multipartite systems of arbitrary dimensions. The CCD was also
used in \cite{confraLAA}, \cite{confraMCSS} to characterize the
input-output equivalent models of networks spin $\frac{1}{2}$, in
a problem motivated by parameter identification for spin
Hamiltonians. Generalizations of these results for networks of
spins of any value, in view of the results presented here, will be
given in a forthcoming paper \cite{confranew}.

\section{Relation between Cartan decompositions and symmetries}
\label{RBC}

The results of \cite{BBO1}, \cite{BBO2} associate to the
Concurrence Canonical Decomposition a time reversal symmetry. In
particular, there is a relation between the involution $\theta$
corresponding to
 the CCD and the time reversal symmetry $\Theta$ in (\ref{th})
 (with $N=n$ the number of spin assumed all equal to $\frac{1}{2}$).
This
 relation is given by
 \be{relat}
\theta(A)=\Theta A \Theta^{-1}, \quad \forall A \in su(2^n),
 \ee
where the right hand side needs to be interpreted as composition
of operators. It is also easily seen, using only the fact that the
time reversal symmetry is antiunitary and the general formula
(\ref{gottfried}), that, if $\bar \theta$ is the time reversal
symmetry on observables $iA$, we have \be{tiv} \bar \theta (iA) :=
\Theta iA \Theta^{-1}=-i\Theta A \Theta^{-1}=-i\theta(A).\ee

This rises the question of whether there is in general a one to
one correspondence between symmetries $\Theta$, $\bar \theta$
(\ref{transA}), and Cartan involutions $\theta$ and therefore
Cartan decompositions. Also the question arises on whether formula
(see (\ref{tiv})) \be{tivcond} \bar \theta(iA)=-i\theta(A),
\forall A \in u(n)\ee is always valid. We shall investigate these
issues in this section. We shall see that only a particular class
of symmetries, which we call {\it Cartan symmetries} give rise to
Cartan involutions.

\bd{Cartsym} A symmetry $\Theta$ is called a Cartan symmetry if
and only if $\Theta^2$ is equal to the identity up to a phase
factor. \ed

Cartan symmetries have the property that applied two times to any
state return the physical state unchanged. For example the time
reversal symmetry and the parity (\ref{parity}) are Cartan
symmetries while the space translation symmetry (\ref{space}) is
not a Cartan symmetry.

Whether or not a symmetry is a Cartan symmetry can be verified
once we have its representation in a given basis i.e. (cf.
(\ref{gottfried})) \be{acta} \Theta |\alpha> = X K |\alpha>, \ee
where $X$ is unitary and $K$ is the identity if $\Theta$ is a
unitary symmetry and is the conjugation of all the components of
$|\alpha>$ if $\Theta$ is antiunitary. $\Theta$ is a Cartan
symmetry if and only if \be{CScond} X \bar X=e^{i \phi} I_{n
\times n}, \ee for some $\phi \in \RR$ in the antiunitary case and
$X^2=e^{i\phi} I_{n \times n}$ for some $\phi \in \RR$ in the
unitary case.  This is clearly of the particular orthonormal basis
chosen. If $\Theta$ is antiunitary and $T$ is a unitary
transformation which transforms one orthonormal basis into another
and $XK$ describes the action of the symmetry in one basis then
$TX\bar T^* K$ describes the action of the symmetry in the new
basis. It is easily seen that if $X$ satisfies (\ref{CScond}) so
does $TX\bar T^*$ and an analogous fact holds for unitary
symmetries.

Generalizing the approach in \cite{BBO1} \cite{BBO2} we now give
the following definition.

\bd{induced} The {\it transformation induced by a symmetry}
$\Theta$ on $su(n)$ is defined as \be{transAbis} \theta(A):=\Theta
A \Theta^{-1}. \ee \ed Notice this definition is analogous to the
one of symmetries $\bar \theta$ on observables (\ref{transA})
which we repeat here with different notations: \be{transAtris}
\bar \theta(iA):=\Theta iA \Theta^{-1}, \qquad \forall A \in
su(n). \ee

In order to give an expression of the induced transformation in a
given basis, we consider the antiunitary and the unitary case
separately. In the antiunitary case, if $K$ is the conjugation
$\Theta=XK$,  $\Theta^{-1}=\bar X^* K=K X^*$, $\theta$ which gives
\be{matform1} \theta(A)=X \bar A X^*. \ee Analogously, one obtains
\be{matform2} \theta(A)=XAX^*, \ee in the unitary case.


\bt{SymvsInv} The transformation $\theta$ on $su(n)$ induced by a
symmetry ($\Theta$, $\bar \theta$) is a Cartan involution if and
only if ($\Theta$, $\bar \theta$) is a Cartan symmetry. Moreover,
if ($\Theta$, $\bar \theta$) is antiunitary, we have $\forall A
\in su(n)$, \be{corre1} \bar \theta(iA) := -i \theta(A). \ee
Moreover, if ($\Theta$, $\bar \theta$) is unitary, we have
$\forall A \in su(n)$, \be{corre2} \bar \theta(iA) := i \theta(A).
\ee \et

\bpr It is easily verified that $\theta$ defined in
(\ref{matform1}) or (\ref{matform2}) is a homomorphism.  Moreover,
assume $\Theta$ is a Cartan symmetry. Then we calculate (in the
antiunitary case and analogously in the unitary case)
\be{calculat} \theta^2(A)=X (\bar{X \bar A X^*}) X^*=X \bar X A
\bar X^* X^*=A,  \ee where in the last equality we have used the
fact that $\Theta$ is a Cartan symmetry. Therefore the associated
$\theta$ is a Cartan involution.

Conversely consider a Cartan involution $\theta$ on $su(n)$,
induced by a symmetry $\Theta$.  Then we want to show that
$\Theta$ is  a Cartan symmetry.

Since  $\theta$ must be of the type {\bf AI}, {\bf AII} or {\bf
AIII}, we must be able to write it as (\ref{AIinvo}),
(\ref{AIIinvo}) or (\ref{AIIIinvo}) up to conjugacy. In particular
there exists a unitary $T$ such that (case {\bf AI}) \be{AIbis}
\theta(B)=T \bar T^* \bar B \bar T T^*, \forall B \in su(n),  \ee
 or such that (case {\bf AII})
 \be{AIIbis} \theta(B)=T J \bar T^* \bar B \bar T J^{-1}
 T^*,
 \forall B \in su(n),
\ee or such that  (case {\bf AIII})
 \be{AIIIbis} \theta(B)=T
I_{p,q} T^* B T I_{p,q} T^*, \ee in the $AIII$ case \footnote{In
the case AI in appropriate coordinates the involution is equal to
 conjugation. If $T$ is the matrix that makes the change of
 coordinates,
 every $B \in su(n)$ can be written as $B=TAT^*$ for a unique $A$ in $su(n)$
 and therefore $A=T^*BT$. Now $\theta_I(B)=T\bar A T^*$ and replacing
 $A=T^*BT$, one obtains (\ref{AIbis}). The other cases are analogous.}
We take $\Theta$ in the cases {\bf AI}, {\bf AII} and {\bf AIII}
given  by (cf. (\ref{matform1}) and (\ref{matform2})) \be{AItris}
\Theta=T\bar T^* K,\ee \be{AIItris} \Theta=T J \bar T^* K,\ee and
\be{AIIItris} \Theta=TI_{p,q} T^*,\ee respectively. It is easily
verified that these are all Cartan symmetries, i.e. $X \bar X=I$
with $X= T \bar T^*$, $X=TJ\bar T^*$ and $X^2=I_{n \times n}$ with
$X=TI_{p,q}T^*$. Moreover the choice is unique, up to a phase
factor which does not change the property of the symmetry of being
a Cartan symmetry, as the set of matrices $su(n)$ is an
irreducible set of skew-Hermitian operators.   This concludes the
proof of the theorem. \epr

\br{reu1} The theorem could have been stated in a somewhat
stronger form. In fact, the proof shows not only that the symmetry
corresponding to a Cartan involution is a Cartan symmetry by also
that it exists and is unique up to a phase factor. Therefore there
is a one to one correspondence given by (\ref{transAbis})
(\ref{transAtris}) between Cartan symmetries and Cartan
involutions and therefore decompositions. \er

\br{AUNvsUN} It follows from the proof of the theorem that
antiunitary Cartan symmetries correspond  to Cartan involutions of
the type {\bf AI}  and {\bf AII} while unitary ones give rise to
Cartan involutions of type {\bf AIII.} \er


\section{Dual structures of $u(n)$ and $iu(n)$;
Commutation and anticommutation relations} \label{DSO}
 In this
section, we study the dual structure of the Lie algebra $u(n)$ of
skew-Hermitian matrices and the Jordan algebra $iu(n)$ of
Hermitian matrices equipped with the anticommutator operation. We
shall see that to a Cartan decomposition of $u(n)$ there
correspond a decomposition of $iu(n)$ which we also call `Cartan'
where the role of the subspaces are possibly reversed. This
correspondence is crucial in the development of general
decompositions for multipartite systems developed in the next
section. The situation is somehow different if we consider
decompositions of the type {\bf AI} and {\bf AII} and if we
consider decompositions of the type {\bf AIII}. Therefore we shall
consider the two cases separately. Only the case {\bf AI} and
{\bf AII} will in fact be used in the next section.


Consider a  Cartan decomposition of $su(n)$ (\ref{Cart}),
(\ref{comm1}), (\ref{comm2}), (\ref{comm3}) of the type {\bf AI}
or {\bf AII}, its corresponding Cartan involution $\theta$ and
Cartan symmetry $\bar \theta$ related through (\ref{corre1}). This
decomposition naturally extends to a decomposition of $u(n)$ by
replacing $\cal P$ with ${\cal P} \oplus span \{i I_{n \times n}
\}$. We shall denote this subspace, with some abuse of notation,
again by $\cal P$. So that \be{Cartun} u(n)={\cal K} \oplus {\cal
P}, \ee ${\cal P}={\cal K}^\perp$, where the orthogonal complement
is now taken in $u(n)$, and the commutation relation
(\ref{comm1}), (\ref{comm2}) and (\ref{comm3}) also holds, within
$u(n)$. The Cartan involution $\theta$ of the type {\bf AI} and
{\bf AII,} is  naturally extended to $u(n)$ and $span \{ i I_{n
\times n} \}$ will belong to the $-1$ eigenspace of $\theta$ so
that the new definition of $\cal P$ is consistent with the fact
that $\cal P$ is the $-1$ eigenspace of $\theta$. The
corresponding symmetry on $iu(n)$ will be given by (\ref{corre1})
or equivalently by (\ref{transAtris}).

Consider now $iu(n)$\footnote{From now on in this section  all the
matrices are assumed skew-Hermitian so that matrices multiplied by
$i$ are Hermitian. An exception is in Remark \ref{DD} below where
the $H$'s denote Hermitian operators.} which has a structure of a
Jordan algebra when equipped with the anti-commutator operation
\be{anticom} \{iA, iB \}:=(iA) (iB)+(iB)(iA).  \ee Associated to a
Cartan decomposition of $u(n)$ (\ref{Cartun}) is a decomposition
of $iu(n)$, which we also call Cartan decomposition, given by
\be{Cartiun} iu(n)= i{\cal K} \oplus i{\cal P}. \ee Moreover it
follows from (\ref{transAtris}) that $\bar \theta$ is a
homomorphism on the Jordan algebra
 $iu(n)$. It is in fact an involution as $\bar \theta^2$ is equal to the
 identity map. It follows from (\ref{corre1}) that $i \cal P$ and $i
 \cal K$ are, respectively, the $+1$ and $-1$ eigenspaces of $\bar
 \theta$ and therefore we have
 \be{Cartanti}
\{i {\cal P}, i {\cal P} \} \subseteq i {\cal P}, \qquad \{ i
{\cal P}, i{\cal K} \} \subseteq i{\cal K}, \qquad \{i {\cal K}, i
{\cal K} \} \subseteq i{\cal P}. \ee So the roles of the subspaces
$\cal K$ and $\cal P$ are somehow reversed when going from $u(n)$
to $iu(n)$.


In the case of a Cartan decomposition of the type {\bf AIII} of
$su(n)$, the construction is similar. In this case, we extend the
Cartan decomposition to $u(n)$ by incorporating $span \{i I_{n
\times n}\}$ into $\cal K$ rather than
 into $\cal P$. The commutation relations (\ref{comm1})
 (\ref{comm2}) and (\ref{comm3}) are still valid with this modified
 definition. The induced decomposition on $iu(n)$ given in
 (\ref{Cartiun}) is such that $i{\cal K}$ and $i {\cal P}$ are the
 $+1$ and $-1$ eigenspaces of the involution $\bar \theta$. This
 follows from the correspondence between $\theta$ and $\bar \theta$
 which in this case is given by (\ref{corre2}). We have
\be{inclusionsAIII} \{ i {\cal K}, i {\cal K} \} \subseteq i {\cal
K}, \qquad \{ i{\cal P}, i{\cal K} \} \subseteq i {\cal P}, \qquad
\{ i {\cal P}, i {\cal P} \} \subseteq i {\cal K}.  \ee


\br{DD} {(Decomposition of dynamics)} It was pointed out in
(\cite{BBO2}) that every evolution of a finite dimensional quantum
system $U:=e^{i H}$ can be decomposed as \be{deco111}
e^{iH}=e^{iH_a}e^{iH_s},  \ee where the Hamiltonian $H_s$ is
symmetric with respect to time reversal symmetry and the
Hamiltonian $H_a$ is antisymmetric (cf.
(\ref{symsym1})-(\ref{symsym3})) with respect to time symmetry,
i.e. $\bar \theta (H_s)=H_s$ and $\bar \theta (H_a)=-H_a$.
Therefore every evolution can be decomposed into a time symmetric
one and a time antisymmetric one. In view of the above treatment
such a decomposition can be extended to any Cartan symmetry. For
Cartan symmetries of the type {\bf AI} and {\bf AII}, $iH_a$ and
$iH_s$ are in the $\cal K$ and $\cal P$ subspace of the associated
decomposition so that the decomposition (\ref{deco111}) is Cartan
decomposition (\ref{elem}). To this purpose also notice that the
$\cal K$ Lie algebras  in all the three types of Cartan
decompositions correspond to semisimple compact Lie groups so that
the exponential map is surjective \cite{Helgason}. The same
argument can be repeated for Cartan symmetries of the type {\bf
AIII} with the only change that this time $iH_s \in{\cal K}$ and
$i H_a \in {\cal P}$. \er

\section{Cartan decompositions for  multipartite systems in arbitrary dimensions;
Decompositions of the odd-even type}

\label{EMS}

In this section we shall generalize the Concurrence Canonical
Decomposition to the general case i.e. to the case of a
multipartite system consisting of { any} number of  quantum
systems of {\it any} dimension. We shall call the general
decomposition a {\it decomposition of the  even-odd type} because
the two subspaces in the Cartan decomposition consist of elements
which are tensor products of an odd or even number of elements in
appropriate subspaces. In doing this, we shall make use of the
correspondence between decompositions in $u(n)$ and decompositions
in $iu(n)$ described in the previous section. In particular, we
shall consider decompositions of $u(n)$-$iu(n)$ associated to
antiunitary Cartan symmetries. To this correspond Cartan
decompositions and involutions which have the property to extend
to Cartan decompositions and involutions  for multipartite systems
as we shall now describe.


Consider a multipartite quantum system composed of $N$ quantum
systems of dimensions $n_1$, $n_2$,...,$n_N$ and with Hilbert
spaces ${\cal H}_1,$...,${\cal H}_N$. The space of skew-Hermitian
(Hermitian) operators acting on the space ${\cal H}_j$,
$j=1,...,N$ is $u(n_j)$ ($iu(n_j)$). The space of skew-Hermitian
(Hermitian) operators acting on the total Hilbert space ${\cal
H}_{TOT}:={\cal H}_1 \otimes {\cal H}_2 \otimes \cdot \cdot \cdot
\otimes {\cal H_N}$ is $u(n_1n_2\cdot \cdot \cdot n_N)$
(i$u(n_1n_2\cdot \cdot \cdot n_N)$). Consider now Cartan
decompositions of $u(n_j)$, not necessarily all of the same type
but all of the type {\bf AI} or {\bf AII}, \be{CartNj}
u(n_j)={\cal K}_j \oplus {\cal P}_j, \ee and the corresponding
decompositions for $iu(n_j)$ \be{CartNjHermit} iu(n_j)=i{\cal K}_j
\oplus i{\cal P}_j, \ee satisfying, with obvious modification of
the notations, the commutation relations (\ref{comm1}),
(\ref{comm2}), (\ref{comm3}) and anticommutation relations
(\ref{Cartanti}). Let us denote by $\sigma_j$ a generic element of
an orthonormal basis in $i {\cal K}_j$ which is an Hermitian
matrix. Also let us denote by $S_j$ a generic element of an
orthonormal basis in $i {\cal P}_j$ which is also an Hermitian
matrix. An orthonormal basis in $iu(n_1n_2\cdot \cdot \cdot n_N)$
is given by tensor products of the form \be{Formatensprod} F:= T_1
\otimes T_2 \otimes \cdot \cdot \cdot \otimes T_N, \ee  where
$T_{j}=\sigma_{j}$ or $T_{j}=S_{j}$, with all the possible
combinations of $\sigma$'s and $S$'s in the $N$ places. We define
${\cal I}_o$ ($ {\cal I}_e$) the subspace of $iu(n_1n_2\cdot \cdot
\cdot n_N)$ spanned by tensor products which display an odd (even)
number of elements $\sigma$, so that we write \be{decomul}
iu(n_1n_2\cdot \cdot \cdot n_N)={\cal I}_o \oplus {\cal I}_e. \ee
We shall call this decomposition,  along with the corresponding
decomposition of $u(n_1n_2 \cdot \cdot \cdot n_N)$ \be{decomulSH}
u(n_1 n_2 \cdot \cdot \cdot n_N)=i{\cal I}_{o} \oplus i {\cal
I}_e, \ee a decomposition of the {\it odd-even type}. We have the
following result \bt{Main} The decomposition of the odd-even type
(\ref{decomul}) (\ref{decomulSH})  is a Cartan decomposition which
is associated to a antiunitary Cartan symmetry, i.e. \be{CDmult}
[i{\cal I}_o,i{\cal I}_o] \subseteq i{\cal I}_o, \qquad [i{\cal
I}_o,i{\cal I}_e] \subseteq i{\cal I}_e, \qquad [i{\cal
I}_e,i{\cal I}_e] \subseteq i{\cal I}_o,  \ee \be{CDmultH} \{{\cal
I}_o,{\cal I}_o\} \subseteq {\cal I}_e, \qquad \{{\cal I}_o,{\cal
I}_e\} \subseteq {\cal I}_o, \qquad \{ {\cal I}_e, {\cal I}_e \}
\subseteq {\cal I}_e \ee \et \bpr The proof is by induction on the
number of systems $N$. For $N=1$, ${\cal I}_o=i{\cal K}_1$ and
${\cal I}_e=i {\cal P}$ so that the commutation and
anticommutation relations (\ref{CDmult}) and (\ref{CDmultH}) are
the same as (\ref{comm1})-(\ref{comm3}) and (\ref{Cartanti}),
respectively. Assuming now (\ref{CDmult}) and (\ref{CDmultH}) true
for every number of subspaces strictly less then $N$ we can verify
(\ref{CDmult})  for $N$ by using the formula \be{formula1} [A
\otimes B, C \otimes D]= \frac{1}{2}([A,C] \otimes \{B,D\} +
\{A,C\} \otimes [B,D]), \ee and considering all the subcases. For
example, to show the first one of (\ref{CDmult}) one considers the
four cases,  by indicating with the superscript the number of
factors in the tensor products:

C1: $\{$ $A \in {\cal I}_{o}^{N-1}$, $B \in i {\cal I}_{e}^1$, $C
\in {\cal I}_{o}^{N-1}$, $D \in i{\cal I}_{e}^1$ $\}$

C2: $\{$ $A \in {\cal I}_{o}^{N-1}$, $B \in i {\cal I}_{e}^1$, $C
\in {\cal I}_{e}^{N-1}$, $D \in i{\cal I}_{o}^1$ $\}$

 C3: $\{$ $A \in {\cal I}_{e}^{N-1}$, $B
\in i {\cal I}_{o}^1$, $C \in {\cal I}_{o}^{N-1}$, $D \in i{\cal
I}_{e}^1$ $\}$

C4: $\{$ $A \in {\cal I}_{e}^{N-1}$, $B \in i {\cal I}_{o}^1$, $C
\in {\cal I}_{e}^{N-1}$, $D \in i{\cal I}_{o}^1$ $\}$

 Analogously, one can verify (\ref{CDmultH}) by using induction
along with the formula \be{formula2} \{A \otimes B, C \otimes D
\}= \frac{1}{2} ([A,C] \otimes [B,D] + \{A,C \} \otimes \{B, D\}
). \ee
 \epr

Associated to a decomposition of the odd-even type is a Cartan
involution on $u(n_1n_2\cdot \cdot \cdot n_N)$, $\theta^{TOT}$,
and the corresponding Cartan symmetry  on the space
$iu(n_1n_2\cdot \cdot \cdot n_N)$, $\bar \theta^{TOT}$. If
$\theta_1,...,\theta_N$ and $\bar \theta_1,...,\bar \theta_N$ are
the Cartan involutions and symmetries  associated to the
$1,2,...,N$-th decomposition, $\theta^{TOT}$ and $\bar
\theta^{TOT}$ can be described as follows.

Let $A$ be  an element of the orthonormal basis of $i{\cal I}_o$,
i.e. it can be written as \be{Adefin} A=
T_{1}\otimes\cdots\otimes(iT_{k}) \otimes \cdot \cdot \cdot
\otimes T_N,   \ee  where $T_{j}=\sigma_{j}$ or $T_{j}=S_{j}$,
with an odd number of $\sigma$'s. Then \be{thetaTOT}
\theta^{TOT}(A)=\bar{\theta}_1(T_{1}) \otimes\cdots\otimes
\theta_{k}(iT_{k})
 \otimes\cdots\otimes \bar \theta_{N}(T_N)= \pm A, \ee
 since $\bar{\theta}_{j}(\sigma_{j})=-\sigma_{j}$,
 $\bar{\theta}_{j}(S_{j})=S_{j}$, ${\theta}_{k}(i\sigma_{k})=i\sigma_{k}$,
 and ${\theta}_{k}(iS_{k})=-iS_{k}$.

In general an element of the orthonormal basis in $u(n_1 n_2 \cdot
\cdot \cdot n_N)$ is a tensor product of $\sigma$ and $S$
elements, with $i$ multiplying one of the elements. $\theta^{TOT}$
is obtained by applying $\bar \theta_j$ in all the positions $j$
without $i$ and $\theta_k$ in the $k-$th position where there is
the factor $i$. If in the $k-$th position there is a factor of the
type $\sigma$ this gives a $+1i\sigma$ factor when transformed. In
the remaining terms, all the factors $S$ are transformed into $S$
while factors $\sigma$ give other factors of the type $\sigma$ and
a collective factor $(-1)^{p-1}$. Here $p$ is the total number of
$\sigma$'s and this is $1$ if $p$ is odd and $-1$ if $p$ is even,
so that $\theta^{TOT}(A)=A$ in one case and $\theta^{TOT}(A)=-A$
in the other case.

Analogously, one can treat the case where in the $k-$th position
there is a factor of the type $iS$. With a similar argument, one
shows that $\bar \theta^{TOT}$ can be defined on tensor products
by applying $\bar \theta_j$ in every $j-$th position which clearly
gives a factor $(-1)^p$ where $p$ is the number of factors
$\sigma$. This shows $\theta^{TOT}$ and $\bar \theta^{TOT}$ are
the involution and symmetry associated
 with the odd-even Cartan decomposition.

 An alternative treatment could have been to first define
 the involutions and symmetries and then to  obtain the
 decomposition (\ref{decomul})-(\ref{CDmultH}) in terms of
 eiegenspaces of these homomorphisms.

 \br{CCDdestroyed}
The Concurrence Canonical Decomposition is obtained as a special
case of the odd-even decomposition when the $N$ systems are all
two level systems and the Cartan decomposition chosen on each of
them is of the type {\bf AII}. This gives ${\cal K}=su(2)=sp(1)$
and ${\cal P}=\{0\}$ in the decomposition of $su(2)$ (\ref{Cart}).
It corresponds to a time reversal symmetry ((\ref{th}) with $N=1$
and spin $\frac{1}{2}$) which is indeed a Cartan symmetry. Notice
that $n=2$ is the only case where in the Cartan decompositions
{\bf AI}, {\bf AII} and {\bf AIII}, we can take $\cal K$ equal to
the whole Lie algebra $su(n)$.  This fact makes it difficult, in
higher dimensions,  to obtain
 natural decompositions of dynamics into local and entangling parts
 as it was done for the 2-qubits case for example in \cite{Zhang}.
 \er

\section{The nature of the odd-even decomposition}
\label{TNO}

 It follows (for instance) from formulas (\ref{CDmult})
(\ref{CDmultH}) that the odd-even decomposition is a decomposition
of the type {\bf AI} or {\bf AII}, namely a decomposition
corresponding to a Cartan symmetry. It is interesting to know how
the choice of the single decompositions on the various subsystems
determines whether the odd-even decomposition is of the type {\bf
AI} or {\bf AII}. One reason for that is that one may want to
further decompose the Lie algebra $i{\cal I}_o$ and therefore
would like to know its nature. For example, it was shown in
\cite{BBO2} that the Concurrence Canonical Decomposition is {\bf
AI} in the case of even number of qubit subsystems  and {\bf AII}
in the case of odd qubits. In our notation $i{\cal I}_o$ is
(conjugate to) $so(2^N)$ for $N$ even and $sp(2^{N-1})$ for $N$
odd.


In general this information can be obtained by a simple count of
the dimensions. Recall that in a {\bf AI}  decomposition of $u(n)$
the dimension of the Lie algebra $\cal K$ in (\ref{Cartun}) is the
dimension of $so(n)$ i.e. $d_I:=\frac{n(n-1)}{2}$ while in a {\bf
AII}  decomposition of $u(n)$ the dimension of the Lie algebra
$\cal K$ is the dimension of $sp(\frac{n}{2})$ i.e.
$d_{II}:=\frac{n(n+1)}{2}$. These numbers are never the same and
therefore they uniquely identify the type of decomposition
obtained. We have the following result.

\bt{dim} Consider an odd-even decomposition on $N$ subsystem
obtained  by performing  {\bf AII} decompositions on  $r$
subsystems and  {\bf AI} decompositions on $N-r$ subsystems. Then
the resulting decomposition is of type {\bf AII} if $r$ is odd and
of type {\bf AI} if $r$ is even. \et \bpr The proof is by
induction on $N$. If $N$ is equal to $1$ the result is obvious.
Consider now $N$ subsystems and consider first the case $r$ odd.
Assume, without loss of generality, that an {\bf AII}
decomposition is performed on the $N-$th subsystem. Let $n_1$
($n_2$) denote the dimension of the vector composed of the first
$N-1$ systems (of the $N-$th subsystem). By the inductive
assumption the odd-even decomposition on  the first $N-1$ system
is of the {\bf AI} type. Therefore there are
$\frac{n_1(n_1-1)}{2}$ elements of the odd type i.e. tensor
products containing an odd number of $\sigma$ matrices spanning
the associated subalgebra $\cal K$. A remaining orthonormal set of
$n^2_1- \frac{n_1(n_1-1)}{2}=\frac{n_1(n_1+1)}{2}$ even type
elements  span the orthogonal complement in $u(n_1)$. The basis
for the Lie algebra $\cal K$ for the system composed of all the
$N$ subsystems is obtained by tensor products of the
$\frac{n_1(n_1-1)}{2}$ odd elements with the
$n_2^2-\frac{n_2(n_2+1)}{2}=\frac{n_2(n_2-1)}{2}$ even type
elements on the $N-$th subsystem or by products of the
$\frac{n_1(n_1+1)}{2}$ even elements with the
$\frac{n_2(n_2+1)}{2}$ even type elements on the $N-$th subsystem.
The dimension of the $\cal K$ Lie algebra in the resulting
odd-even decomposition is therefore \be{dimension}
\frac{n_1(n_1-1)}{2} \times \frac{n_2(n_2-1)}{2}
+\frac{n_1(n_1+1)}{2} \times \frac{n_2(n_2+1)}{2}=\frac{n_1n_2(n_1
n_2+1)}{2}, \ee which indicates a decomposition of the type {\bf
AII} for the total system which has dimension $n_1n_2$. An
analogous reasoning proves the case where $r$ is even. \epr

The result of \cite{BBO1} is obtained as a special case of the
above theorem as in the case treated all the decompositions
applied are of type {\bf AII} and therefore the Concurrence
Canonical Decomposition is of type {\bf AI} on an even number of
subsystems and of type {\bf AII} on an odd number.


\vs

{\bf Acknowledgment} D. D'Alessandro's research was supported by
NSF under Career Grant ECS-0237925.

\end{document}